\documentclass[10pt,twocolumn]{article}
\usepackage{makecell}
\usepackage[utf8]{inputenc}
\usepackage[T1]{fontenc}
\usepackage{subcaption}
\usepackage{amsmath}
\usepackage{tabularx}
\usepackage{lipsum}
\usepackage{mathtools}
\usepackage{cuted}
\usepackage{flushend}
\usepackage{amsmath,amsthm,verbatim,amssymb,amsfonts,amscd, graphicx}
\usepackage{graphics}
\usepackage[colorlinks,linkcolor=blue,citecolor=blue]{hyperref}
\usepackage{helvet}  
\usepackage{courier}  
\usepackage{url}  
\usepackage{graphicx}  
\usepackage[numbers, sort&compress]{natbib}
\usepackage{cuted}
\usepackage{flushend}

\begin{document}

\title{Dynamical activity universally bounds precision of response in Markovian nonequilibrium systems}

\author{Kangqiao Liu$^{1, *}$, Jie Gu $^{2, +}$\\
{\small
\textit{$^1$School of Science, Key Laboratory of High Performance Scientific Computation,}}\\ {\small\textit{Xihua University, Chengdu 610039, China}}\\
{\small\textit{$^2$Chengdu Academy of Educational Sciences, Chengdu 610036, China}}\\
{\small{$^*$kqliu@xhu.edu.cn}}\\
{\small{$^+$jiegu1989@gmail.com}}}

\maketitle


\begin{abstract}
The exploration of far-from-equilibrium systems has been at the forefront of nonequilibrium thermodynamics, with a particular focus on understanding the fluctuations and response of thermodynamic systems to external perturbations. In this study, we introduce a universal response kinetic uncertainty relation, which provides a fundamental trade-off between the precision of response for generic observables and dynamical activity in Markovian nonequilibrium systems. We demonstrate the practical applicability and tightness of the derived bound through illustrative examples. Our results are applicable to a broad spectrum of Markov jump processes, ranging from currents to non-current variables, from steady states to time-dependent driving, from continuous time to discrete time, and including Maxwell's demon or absolute irreversibility.  Our findings not only enhance the theoretical foundation of stochastic thermodynamics but also may hold potential implications for far-from-equilibrium biochemical processes.
\end{abstract}

\thispagestyle{empty}

\section*{Introduction}
The out-of-equilibrium regime, characterized by systems subjected to external driving forces or gradients, presents a particularly intriguing domain for stochastic thermodynamics \cite{seifert2012,peliti2021,shiraishi2023}. A fundamental and remarkable principle recently emerged in this regime is the thermodynamic uncertainty relation (TUR) \cite{barato2015,gingrich2016}
\begin{equation}\label{eq:TUR}
	\frac{ {J}^{2}}{\langle\langle{J}\rangle\rangle} \leq \frac{ {\sigma}}{2},
\end{equation}
where ${J}$ is the ensemble-averaged time-integrated nonequilibrium current, $\langle \langle \cdot \rangle \rangle$ represents the variance of a random variable, and $\sigma$ is the total entropy production (EP) characterizing thermodynamic dissipation defined explicitly in Eq.~\eqref{eq: EP}. The conventional TUR \eqref{eq:TUR} has been proved true for the nonequilibrium steady state (NESS) of Markov processes \cite{gingrich2016}, but not tight for systems with a discrete state space \cite{vo2022}. This indicates that the EP alone may not be adequate to bound the fluctuations of the system fully. For systems that are significantly far from equilibrium, it is essential to consider non-dissipative factors as well. A key metric found in this context is the dynamical activity (DA) \cite{maes2006,baiesi2009a,maes2017,maes2018,maes2020}, which represents the average number of state transitions occurred in a stochastic process thus characterizing the kinetics. 
The notion of DA  originates from the development of nonequilibrium phase transitions \cite{Garrahan2007,Lecomte2007,Baiesi2009,Fullerton2013} and has an intimate relationship with thermo-kinetic responses to external perturbations \cite{Baiesi2009,Falasco2016}.
In parallel to the TUR, it has been proved for the time-homogeneous Markov jump process that the precision of a current ${J}$ is also bounded by the kinetic uncertainty relation (KUR) \cite{diterlizzi2019} (see also \cite{pietzonka2016,garrahan2017,chiuchiu2018}),
\begin{equation}\label{eq:KUR}
	\frac{ {J}^{2}}{\langle\langle{J}\rangle\rangle} \leq \mathcal{A},
\end{equation}
where $\mathcal{A}$ is the DA \cite{maes2006,baiesi2009a,maes2017,maes2018,shiraishi2018,maes2020}.  The TUR \eqref{eq:TUR} and the KUR \eqref{eq:KUR} have complementary roles and they can be derived from a unified thermodynamic-kinetic uncertainty relation (TKUR) \cite{vo2022}.

\begin{figure}[t]
    \centering
    \includegraphics[width=\columnwidth]{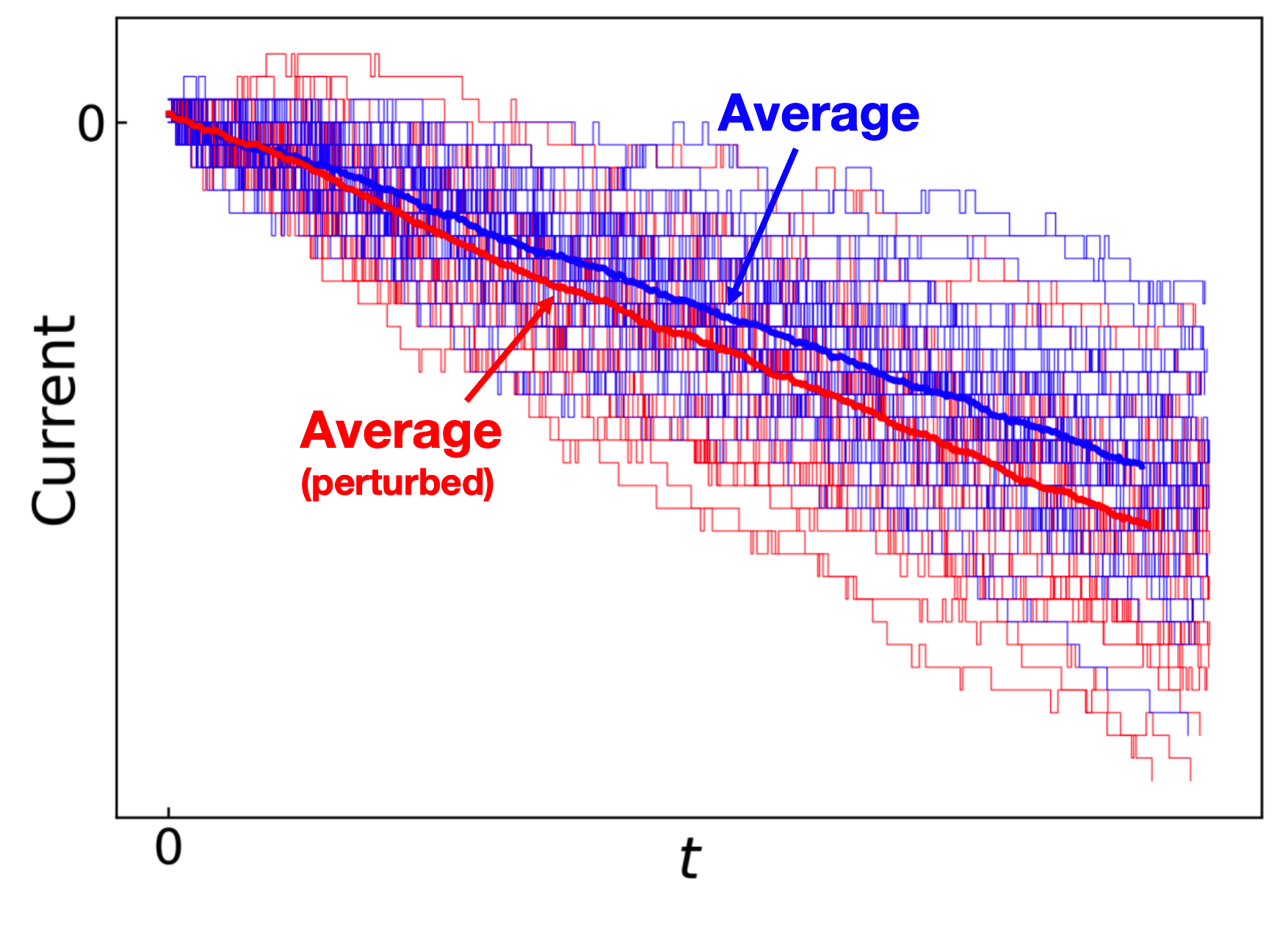}
    \caption{\textbf{Monte Carlo simulation of a time-integrated current $J$ during the relaxation process of time $t$ of a three-state model.} We use blue color to depict the process with a certain value of the control parameter $\theta$ and use red color for a perturbed control parameter $\theta+{\rm d}\theta$. The two thick curves represent the average current and each thin curve is a stochastic trajectory. The deviation of the average values represents ${\rm d}J$ due to a change ${\rm d}\theta$.}
    \label{fig:schematic}
\end{figure}

\begin{table*}[t!]
        \centering
\begin{tabularx}{\textwidth}{l c c c c X}
\hline 
  & NESS & Relaxation & Time-dependent driving & Observable & Equality attainability for discrete-state \\
\hline
 TUR \cite{barato2015,gingrich2016} \eqref{eq:TUR} & yes & no  & no & current & no\\
\hline 
 KUR \cite{pietzonka2016,garrahan2017,chiuchiu2018,diterlizzi2019} \eqref{eq:KUR} & yes & yes & no & generic & no \\
\hline 
R-TUR \cite{ptaszynski2024PRL,aslyamov2024b} \eqref{eq:R-TUR}  & yes  & unknown & unknown & current & unknown\\
\hline 
 R-KUR \eqref{eq:bound} & yes & yes & yes & generic & yes\\
 \hline
\end{tabularx}
\caption{
\label{table}
\textbf{Comparison of the range of applicability for the TUR, the  KUR, the R-TUR, and our R-KUR.} The abbreviation NESS means the nonequilibrium steady state.}
\end{table*}

These relations quantify the trade-offs between intrinsic fluctuations and thermodynamic quantities or system dynamics, providing profound insights into the properties of systems operating out-of-equilibrium. It is also worth noting that the importance of the thermodynamic limitation is more pronounced near equilibrium, whereas the kinetic bound predominantly dictates the precision of currents in the far-from-equilibrium regime \cite{vo2022}. This fact suggests the significance of the DA for systems far away from equilibrium \cite{shiraishi2018,vanvu2023,gu2024}.

Concurrently, the development of finding fundamental principles characterizing nonequilibrium systems' static response to external perturbations is also prosperous \cite{Dechant2020,owen2020,owen2023,fernandesmartins2023,aslyamov2024a,aslyamov2024,zheng2024}  (see Fig. \ref{fig:schematic} for a schematic illustration), apart from the seminal fluctuation-dissipation relation \cite{callen1951,kubo1966a} and its generalizations in the far-from-equilibrium regime \cite{cugliandolo1994,ruelle1998,harada2005a,lippiello2005,speck2006,baiesi2009a,prost2009a,seifert2010,lippiello2014,altaner2016}.
Notably, a response thermodynamic uncertainty relation (R-TUR) valid for the NESS of a homogeneous Markov jump process have been proposed very recently \cite{ptaszynski2024PRL,aslyamov2024b}, expressed as:
\begin{equation}\label{eq:R-TUR}
	\frac{\left(\text{d}_{\theta} {J}\right)^{2}}{\langle\langle{J}\rangle\rangle} \leq \frac{b_{\max }^{2} {\sigma}}{2}.
\end{equation}
Here, $\text{d}_\theta {J}:= \text{d} {J}(\theta)/\text{d} \theta$ is the static response of the current ${J}$ to the perturbation in the control parameter $\theta$ and $b_{\max} = \max_{ij}|\text{d}_\theta B_{ij}|$ is the maximal rate of change in kinetic barriers, with the transition rate parameterized by $\theta$ as
\begin{equation}
\label{eq:para}
    W_{ij}^\theta = e^{B_{ij}(\theta)+S_{ij}/2},
\end{equation}
where the symmetric part $B_{ij}(\theta)$ is the kinetic barrier and the anti-symmetric part $S_{ij}$ is the entropy change associated with the jump from $j$ to $i$. The kinetic barrier can be controlled by, e.g., the enzyme concentration in biochemical processes \cite{wachtel2018} or gate voltages in nanoelectronics \cite{datta1997electronic,nazarov2009quantum}, and the anti-symmetric part corresponds to thermodynamic forces such as chemical potential gradient \cite{seifert2012}. This R-TUR cannot be derived from the conventional TUR \eqref{eq:TUR}, but it instead implies the TUR and offers a tighter bound on the dissipation.

A natural and compelling question arises: Does there exist a counterpart of the R-TUR \eqref{eq:R-TUR} in terms of the DA? In other words, is there a fundamental relationship between the precision of response and the DA? In this work, we answer this question by establishing a rigorous response kinetic uncertainty relation (R-KUR) ---a trade-off between the precision of response of generic observables to arbitrary perturbations and the DA [c.f. inequality \eqref{eq:bound}] as a complementary but more general counterpart of the R-TUR. We illustrate the strength and tightness of our bound with concrete examples. The R-KUR applies to arbitrary Markov jump processes, extending the scope of perturbation from kinetic to arbitrary, from NESS to time-dependent driving, from currents to non-current variables, from continuous-time to discrete jumps, and including processes with Maxwell's demon or unidirectional transitions.

\section*{Results and Discussion}
\subsection*{Setup and main result}
Consider a stochastic Markov jump process with discrete states. The dynamics of the probability distribution is described by the master equation \cite{kampen1992}
\begin{equation}
\label{eq:me}
    \frac{\text{d} {p}_{i}(t)}{\text{d} t}=\sum_{j} W_{ij}(t) p_{j}(t) = \sum_{j,\nu} W_{ij}^{\nu}(t) p_{j}(t),
\end{equation}
where \(p_i(t)\) is the probability of state \(i\) at time $t$, $W_{ij}^{\nu}(t)$, for $i \ne j$, is the time-dependent transition rate from state $j$ to state $i$ via channel $\nu$ (e.g., by coupling with the $\nu$th reservoir), and \(W_{ij}(t) =\sum_\nu W_{ij}^{\nu}(t)\). 
By conservation of probability, the escape rate of leaving a state $i$ is then \(W_{ii} = -\sum_{j(\ne i)} W_{ij}\).
With the master equation, the total EP for an interval $[0,\tau]$ is explicitly given by \cite{Esposito2010,liu2020}
\begin{equation}
    \label{eq: EP}
    \sigma = \int_0^{\tau} {\rm d}t \sum_{i\ne j, \nu}W_{ij}^{\nu}(\theta,t)p_j(t)\ln \frac{W_{ij}^{\nu}(\theta,t)p_j(t)}{W_{ji}^{\nu}(\theta,t)p_i(t)}.
\end{equation}

The transition rate has a general form as $W_{ij}^{\nu} = \exp(B_{ij}^{\nu} + E_j^{\nu} + S_{ij}^{\nu}/2)$, where $B_{ij}^{\nu}$ is the symmetric part, $E_j^{\nu}$ is the vertex part, and $S_{ij}^{\nu}$ is the anti-symmetric contribution \cite{owen2020}. Each of these three contributions can be controlled independently. Different expressions of them correspond to different types of transition rates, such as the Arrhenius type \cite{mcnamara1989}, the phenotypic transition rates \cite{ge2015}, and the phonon spectral type \cite{ren2010}. We generically parameterize the transition rates as $W_{ij}^\nu(\theta,t)$, where $\theta$ is the control parameter. The only requirement is that the control parameter $\theta$ does not explicitly depend on the time $t$.

 As proved in Methods, using the Cram\'er-Rao inequality yields the main result
\begin{equation}
\label{eq:bound}
	\frac{(\partial_\theta {O})^2}{\langle \langle {O} \rangle \rangle} 
	 \le a_{\max}^2 \mathcal{A},
	\end{equation}
where $O$ is an arbitrary observable which is not necessarily a current or an absolute counting variable,
\begin{equation}
    a_{\max} := \mathop{\max_{ 
i\ne j, \nu}}_{t\in [0,\tau]}
|\partial_\theta \ln W_{ij}^\nu(\theta,t)|
\end{equation}
is a generalization of  $b_{\rm max}$ in the R-TUR \eqref{eq:R-TUR}, and $\mathcal{A}$ is the ensemble-averaged time-integrated DA defined explicitly as \cite{shiraishi2018}
\begin{equation}
	\mathcal{A} :=  \int_0^\tau \text{d}t  \sum_{i\ne j} W_{ij}(\theta,t)p_j(t).
\end{equation}

We introduce the sensitivity as $\partial_\theta \ln O := \partial_\theta O / O$ \cite{aslyamov2024a}, and define the relative precision as the ratio of the square of the average $O^2$ to its variance $\langle \langle O \rangle \rangle$ as appeared in the TUR and the KUR. Inequality \eqref{eq:bound} is recast into
\begin{equation}
    (\partial_\theta \ln {O})^2 \cdot \frac{{O}^2}{\langle \langle {O} \rangle \rangle} \le a_{\max}^2 \mathcal{A}.
\end{equation}
It has a clear physical meaning: enhancing sensitivity and precision necessarily entails more frequent jumps. It also indicates that for a given stochastic dynamics (fixed frequency of transitions), high sensitivity to the change in a control parameter implies large fluctuations or low precision.

By comparing inequalities \eqref{eq:R-TUR} and \eqref{eq:bound}, the R-KUR can be regarded as a complementary counterpart of the R-TUR in terms of the DA, yet transcending its limitation on perturbations of the kinetic barrier or time-independent transition rates. In fact, the R-KUR can be applied to arbitrary Markov jump processes as summarized in Table \ref{table} and does not necessitate perturbations to the driving speed or observation time, making it more versatile in applications. We also note that the R-KUR is valid for any process that can be described by the master equation \eqref{eq:me}, without being restricted to thermodynamic systems. 
Thermodynamic interpretation can be attained by introducing local detailed balance conditions for transition rates, namely $W_{ij}^{\nu}\exp(-\beta_{\nu}\omega_j^\nu) = W_{ji}^{\nu}\exp(-\beta_{\nu}\omega_i^\nu)$, where $\beta_{\nu}$ is the inverse temperature of the $\nu$th reservoir and $\omega_i^\nu := \epsilon_i - \mu^{\nu}N_i$ is the grand potential associated with the $\nu$th reservoir, with $\epsilon_i$ being the energy level of the $i$th inner state of the system, $\mu^{\nu}$ the chemical potential of the $\nu$th reservoir, and $N_i$ the number of particles in state $i$ of the system \cite{Esposito2009,Esposito2012}.

\begin{figure}[t!]
\centering
\includegraphics[width=\columnwidth]{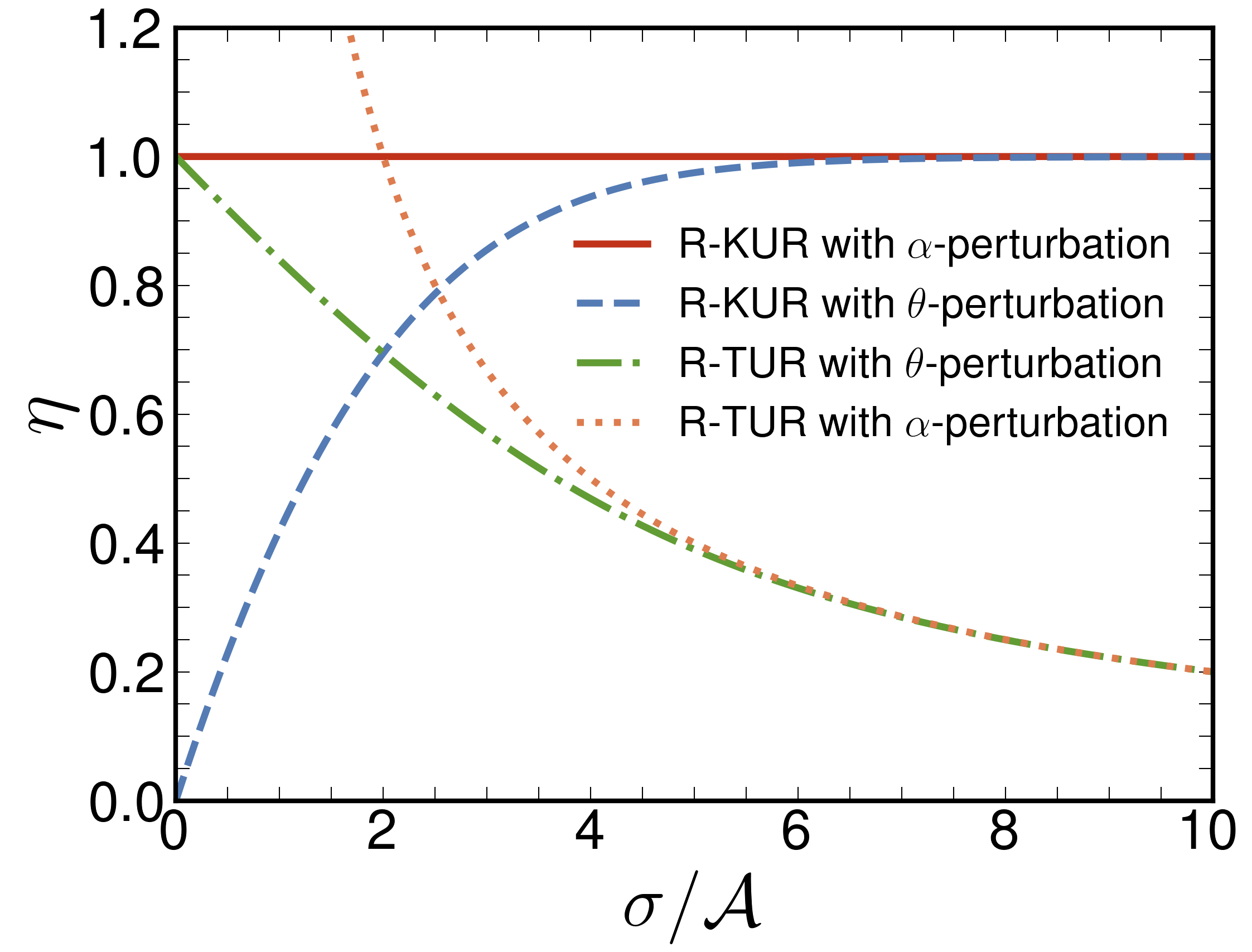}
\caption{
\label{fig:random walk}
\textbf{Numerical simulation of the efficiency of the bound $\eta$ of the R-TURs and the R-KURs as a function of the dissipation-activity ratio $\sigma /\mathcal{A}$ for Example 1.} Green, blue, red, and yellow curves represent the efficiency calculated by inequalities \eqref{eq:RTUR theta}-\eqref{eq:RTUR alpha} about the R-TUR and the R-KUR with perturbation in either the symmetric parameter $\theta$ or the anti-symmetric parameter $\alpha$, respectively.}
\end{figure}

\subsection*{Examples}
\subsubsection*{Example 1: one-dimensional biased random walk}
We illustrate our R-KUR \eqref{eq:bound} in a one-dimensional (1D) biased random walk on $\mathbb{Z}$, which is a successful model describing the dynamics of molecular motors in bio-systems, such as the kinesin moving on cytoskeletal polymer microtubules \cite{julicher1997,Schliwa2003,shin2021}.
It is important to understand its transport properties \cite{Thomas2001,lipowsky2001,klumpp2005,chowdhury2005,Kolomeisky2007,Astumian2010}. 
We can study the response of the transport velocity of a molecular motor to an external perturbation via our bound \eqref{eq:bound}. 

Suppose that a random walker jumps from the position $i$ to the position $i\pm 1$ at an Arrhenius-type rate $k_{\pm} := \exp{(\theta\pm \alpha/2)}$  \cite{mcnamara1989}, with $\theta>0$ being the symmetric Kramer's rate \cite{hanggi1990} controlled by the potential barrier height, and the anti-symmetric part $\alpha>0$ determined by an external potential gradient. The observable is chosen as the current $J$ that flows from the negative direction to the positive direction in the NESS. We first introduce a perturbation in the control parameter $\theta$ while keeping $\alpha$ fixed. As calculated in Methods, we find that the R-TUR \eqref{eq:R-TUR}  becomes exactly the same as the inequality~(4) in the original paper of the TUR \cite{barato2015}. It is explicitly written as
\begin{equation}\label{eq:RTUR theta}
     \frac{(\text{d}_{\theta} {J}(\theta))^2}{\langle \langle {J} \rangle \rangle}  = \tau e^{\theta} \frac{e^{\frac{\alpha}{2}} - e^{-\frac{\alpha}{2}}}{\coth{\frac{\alpha}{2}}} \le \frac{\sigma}{2}.
\end{equation}
Hence, the R-TUR \eqref{eq:R-TUR} indeed holds. The validity of the R-KUR bound \eqref{eq:bound} is checked as well:
\begin{equation}\label{eq:RKUR theta}
    \frac{(\text{d}_{\theta} {J}(\theta))^2}{\langle \langle {J} \rangle \rangle} = \frac{\tau e^{\theta}(e^{\frac{\alpha}{2}} - e^{-\frac{\alpha}{2}})^2}{e^{\frac{\alpha}{2}} + e^{-\frac{\alpha}{2}}} \le \mathcal{A}.
\end{equation}

While the R-TUR \eqref{eq:R-TUR} is limited to the response to changes in the symmetric part of the transition rates, namely $B$ in $W$ rather than $S$, our bound \eqref{eq:bound} does not impose such a restriction.
This greatly enhances the applicability and strength of our result.
Suppose that we are now interested in the response to the change in $\alpha$ rather than $\theta$. 
We first check the R-KUR \eqref{eq:bound}. It yields
\begin{equation}\label{eq:RKUR alpha}
     \frac{(\text{d}_{\alpha} {J}(\alpha))^2}{\langle \langle {J} \rangle \rangle} = \frac{\tau e^{\theta}}{4}  (e^{\frac{\alpha}{2}} + e^{-\frac{\alpha}{2}}) = \frac{\mathcal{A}}{4} = a_{\max}^2\mathcal{A}.
\end{equation}
Surprisingly, the R-KUR attains its equality. We then examine the R-TUR \eqref{eq:R-TUR}. 
For this bound to hold, the following inequality must be satisfied:
\begin{equation}\label{eq:RTUR alpha}
     \frac{(\text{d}_{\alpha} {J}(\alpha))^2}{\langle \langle {J} \rangle \rangle} = \frac{\tau e^{\theta}}{4}  (e^{\frac{\alpha}{2}} + e^{-\frac{\alpha}{2}}) \le \frac{\tau e^{\theta} \alpha}{8}  (e^{\frac{\alpha}{2}} - e^{-\frac{\alpha}{2}}),
\end{equation}
which requires $\alpha > 2.399$. Therefore, the R-TUR is possibly broken while $0<\alpha<2.399$. We perform numerical simulations for the above 4 inequalities in Figure~\ref{fig:random walk} by sampling $\theta$, $\alpha$, and $\tau \in [0,10]$ for $10^5$ data points. We define the efficiency of the bound $\eta:= {(\partial_\theta {O})^2}/{\langle \langle {O} \rangle \rangle} a_{\max}^2 \mathcal{A}$ for the R-KUR and $\eta := 2{(\partial_\theta {O})^2}/{\langle \langle {O} \rangle \rangle} a_{\max}^2 \sigma$ for the R-TUR. The validity of the bound implies $\eta\le 1$; the closer $\eta$ is to 1, the tighter the bound is.

We provide several remarks based on this minimal example. It has been proved that neither the TUR nor the KUR can attain equality in a discrete-state system \cite{vo2022}. The equality can be achieved only for systems with a continuous state space, such as overdamped Langevin systems. On the other hand, the optimized R-TUR remains an inequality, albeit tighter than the conventional TUR, and the condition for equality is elusive \cite{ptaszynski2024PRL}. In this example, it is clear that our R-KUR can achieve equality even in a discrete-state system, implying its tightness compared to previous bounds. We mention that another bound that can attain equality in a discrete-state system is the unified TKUR \cite{vo2022}. Since the equality in the R-KUR holds in this example, it is clear that the R-KUR cannot be implied by the conventional KUR \eqref{eq:KUR}.
Actually, if the transition rates can be parameterized as in Eq. \eqref{eq:para}, it is straightforward to derive the KUR from R-KUR by following a similar procedure in Ref. \cite{ptaszynski2024PRL}.
However, for generically parameterized rates, it remains open whether the R-KUR implies the KUR.

\begin{figure}[t!]
\centering
\includegraphics[width=\columnwidth]{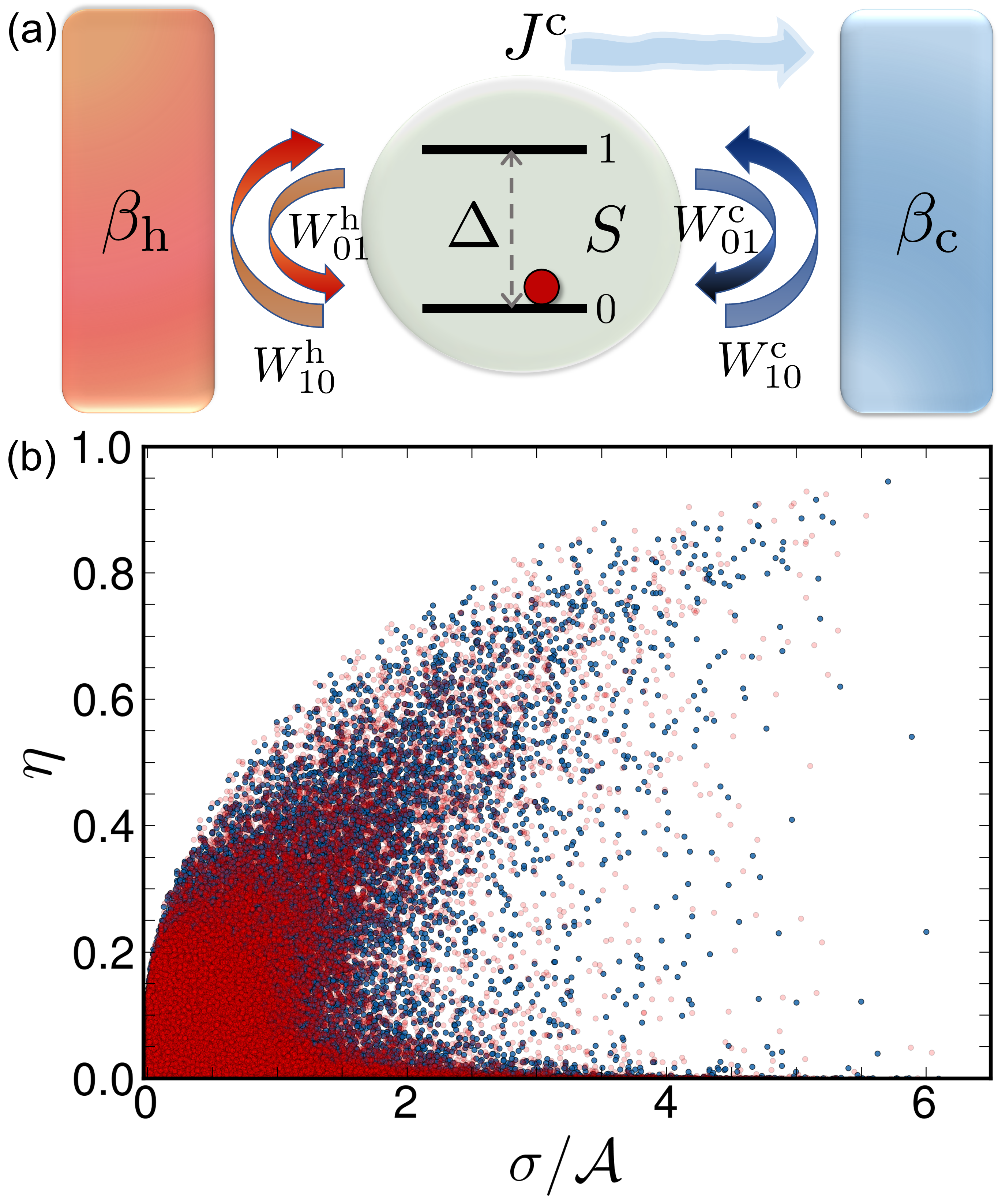}
\caption{
\label{fig}
\textbf{Numerical illustration with a two-level system coupled to two heat baths.} (a) Schematic of a two-level system $S$ coupled with a cold heat bath and a hot bath whose inverse temperatures are denoted as $\beta_{\rm c}$ and $\beta_{\rm h}$, respectively. The energy gap between the 2 levels 0 and 1 is $\Delta$ which can be controlled externally. The transitions can be caused by coupling with either of the baths with 4 transition rates $W_{ij}^{\rm h(c)}$, $i,j\in\{0,1\}$.
(b) Efficiency of the bound $\eta$ versus the dissipation-activity ratio $\sigma /\mathcal{A}$.
Red dots are plotted with a time-integrated current $J^{\rm c}$ flowing into the cold bath; blue dots are plotted with the total number of transitions induced by coupling to the cold bath.
The bound tends to saturation with a large dissipation-activity ratio, indicating that the tightness of the bound increases as the system increasingly deviates from the equilibrium.}
\end{figure}

\subsubsection*{Example 2: two-level system coupled with two heat baths}
The second example involves a two-level system in contact with two baths at different temperatures. 
As schematically depicted in Fig.~\ref{fig}(a), the energy gap is denoted by $\Delta$. We assume the local detailed balance condition $W^\nu_{01}/W^\nu_{10} = e^{\beta_\nu \Delta} $ for each bath $\nu(={\rm c,h})$ with inverse temperature $\beta_\nu$. Suppose that external control can manipulate the energy gap $\Delta$ as a function of time $t$
\begin{equation}
    \Delta(\varepsilon,t) = \varepsilon(C+st+a\sin \omega t). 
\end{equation}
Here only the global amplitude $\varepsilon$ is the parameter to be perturbed while others are kept constant in a specific realization.

We focus on two key observables: one is the time-integrated current flowing into the cold bath $J_{\rm e}[\omega]:= \sum_{i\ne j}(n_{ij}^{\rm c}[\omega] - n_{ji}^{\rm c}[\omega])$; the other is the total number of transitions, or the traffic, induced by coupling to the cold bath $A_{\rm e}^{\rm c}[\omega] := \sum_{i\ne j} n_{ij}^{\rm c}[\omega]$, where $n_{ij}^{\rm c}[\omega]$ counts the total number of transitions from state $j$ to state $i$ via channel c. Their average and variance are calculated by full counting statistics (FCS). 
See Methods for details. In the numerical simulation, we set $W_{10}^{\rm c} = W_{10}^{\rm h} = 1$, $\beta_{\rm h} = 1$. We randomly sampled $\beta_{\rm c}, \varepsilon, C, s, a$, $\omega$ and the time duration $\tau$. Figure \ref{fig}(b) shows a plot of $10^5$ data points for each observable, red for the current and blue for the counting observable, with the efficiency of the bound $\eta := {(\partial_\theta {O})^2}/{\langle \langle {O} \rangle \rangle} a_{\max}^2 \mathcal{A} \le 1$ on the y-axis and the dissipation-activity ratio $\sigma/\mathcal{A}$ on the x-axis.

The bound appears to have the potential to be saturated, particularly in the far-from-equilibrium regime characterized by a large $\sigma/\mathcal{A}$, indicating a tighter bound on the response precision with greater departures from equilibrium, similar to the KUR \cite{vo2022}. We mention that the R-TUR can be broken in this model because the perturbation is not restricted to the symmetric part of the transition rates. Another comment is that no qualitative distinction can be found in the numerical results for different types of  observables because the precision of response is bounded only by the intrinsic kinetics of this process regardless of the structure of the chosen observable. This also supports the claim that our R-KUR is valid for generic observables without any restrictions.




\subsection*{Equality condition}
Example 1 has demonstrated that the R-KUR can attain its equality.
Here we present another sufficient condition of equality.
Consider the steady state of a homogeneous Markov jump process with only one channel.
Inspired by Ref. \cite{shiraishi2021}, we focus on the response around $\theta=0$. The parametrization takes a special form as $W_{ij}(\theta)=W_{ij}\exp(\theta Z_{ij})$ for $i\ne j$ and $W_{jj}(\theta)=-\sum_i W_{ij}\exp( \theta Z_{ij})$, where $Z_{ij}=(W_{ij}p_j^\text{ss}-W_{ji}p_i^\text{ss})/(W_{ij}p_j^\text{ss}+W_{ji}p_i^\text{ss})$ (the superscript ss denotes steady-states) and $W_{ij}$'s are unperturbed transition rates. The empirical observable should take the following form:
\begin{equation}
    \begin{aligned}
        O[\omega] & = \mathcal{J}+\mathcal{I} \\
        & := \sum_{i\ne j} d_{ij} (n_{ij}-n_{ji}) + \sum_i q_i \int_0^\tau \text{d}t \delta_{w(t),i},
    \end{aligned}
\end{equation}
where $d_{ij} :=(W_{ij} p^\text{ss}_j-W_{ji}p^\text{ss}_i)/(W_{ij}p^\text{ss}_j+W_{ji}p^\text{ss}_i)$ and $q_j:=-\sum_i W_{ij}(W_{ij}p^\text{ss}_j-W_{ji}p^\text{ss}_i)/(W_{ij}p^\text{ss}_j+W_{ji}p^\text{ss}_i)$. The first term $\mathcal{J}$ is the time-integrated current where $n_{ij}$ is the number of jumps from state $j$ to $i$.
The second term $\mathcal{I}$ is the so-called time-integrated empirical measure where $w(t)$ is the state of the system at time $t$. It is proved that the average of $O$ vanishes \cite{shiraishi2021}.

The setting above saturates the Cauchy-Schwarz inequality \eqref{eq:cs} and the generalized Cram\'er-Rao inequality since $O[\omega] = \partial_\theta \ln P_\theta[\omega]$ \cite{shiraishi2021}
 (see Methods for definitions of the symbols).
To attain the equality in inequality \eqref{eq:fi}, one further needs the condition that $|\partial_\theta \ln W_{ij}(\theta)|$ is a constant, i.e., $|Z_{ij}|$ is a constant, which is always valid for two-state systems and can be satisfied for systems with more states. 
        
\subsection*{Generalization}
Our main result \eqref{eq:bound} can be extended to many other situations. For example, it is directly applicable to first-passage times \cite{hiura2021} of a Markov jump process. The generalization to the case with multiple control parameters $\Theta:= \{\theta_1, \theta_2, \dots, \theta_M \}$ is also straightforward \cite{dechant2019} given upon that all of these parameters are independent of time $t$ and each other. In the following, we provide further details of the extensions to three important scenarios, including Maxwell's demon, discrete-time Markov chain, and system with unidirectional transitions.


\subsubsection*{Feedback-controlled system}
We consider a system controlled by Maxwell's demon in the context of information thermodynamics \cite{Sagawa2012,Parrondo2015,liu2020}. We investigate a similar process as studied in the TUR for arbitrary initial state \cite{liu2020}. Suppose that the state of the system $\mathbb{S}$ is probed by a meter $\mathbb{M}$. A measurement yields an outcome $m$ with probability $p_m$. The feedback control suddenly changes the transition rate matrix from $\{\mathcal{W}\}_{ij}:= \sum_{\nu} W_{ij}^{\nu}$ to $\mathcal{W}_{m}$ depending on $m$. For simplicity, we assume that the implementation of both the measurement and the feedback control are implemented in infinitesimally short time intervals. 
We also assume homogeneous Markov dynamics. The system then evolves with the new conditional transition matrix $\mathcal{W}_{m}$ for a certain period $\tau$. Finally, the meter is reset to its default state and we repeat this process. We call such a repeated operation a feedback loop. After many loops, the system $\mathbb{S}$ will enter into a stroboscopic steady state $p^{\rm SSS}(t)$ which satisfies $p^{\rm SSS}(t+\tau) = p^{\rm SSS}(t)$, for $t\in [0,\tau]$. It is thus allowed to derive the bound within one feedback loop.

For this feedback-controlled process, the precision of response is bounded from above by 
\begin{equation}
    (\partial_\theta \ln {O})^2 \cdot \frac{{O}^2}{\langle \langle {O} \rangle \rangle} \le \sum_m p_m a_{m,\rm max}^2 \mathcal{A}_m \le a_{\rm max}^2 \mathcal{A},
\end{equation}
where $\mathcal{A}_m := \int_0^{\tau} \text{d}t \sum_{i\ne j, \nu} W_{im,jm}^{\nu}(\theta, t)p_{j|m}(t)$ counts the number of transitions associated with an outcome $m$, $p_{j|m}(t)$ is the conditional probability, and $\mathcal{A}: = \sum_m p_m\mathcal{A}_m$ is the total DA. The optimized constants are defined as $a_{m,\rm max}:= \max_{ 
i\ne j, \nu,t}|\partial_{\theta}\ln W_{im,jm}^{\nu}(\theta,t)|$ and $a_{\rm max} := \max_{m}[a_{m,\rm max}]$. It will be challenging and intriguing to generalize this result to continuous measurement and continuous feedback control \cite{Ribezzi-Crivellari2019,Yada2022}.







\subsubsection*{Discrete-time Markov chain}
The bound \eqref{eq:bound} can be generalized to an inhomogeneous discrete-time Markov chain. Besides mathematics and physics, a discrete-time Markov chain is an important model in many other fields, such as financial markets \cite{Black1973,Ziyin2023}. The updating rule of the probability is described as
\begin{equation}
    P_m(t_i) = \sum_{n,\nu}T_{mn}^{\nu}(\theta,t_{i-1})P_n(t_{i-1}),
\end{equation}
where $P_m(t_i)$ is the probability of the system being in state $m$ at time $t_i$, and $T_{mn}^{\nu}(\theta,t_{i-1})$ is the transition probability from state $n$ to state $m$ through channel $\nu$ at time $t_{i-1}$ with a control parameter $\theta$ satisfying $0 \le T_{mn}^{\nu}(\theta,t_{i-1}) \le 1$ and $\sum_{m,\nu}T_{mn}^{\nu}(\theta,t_{i-1}) = 1$.

We prove in Methods that, for $N$ steps,
\begin{equation}\label{eq:discrete bound}
    (\partial_\theta \ln {O})^2 \cdot \frac{{O}^2}{\langle \langle {O} \rangle \rangle} \le \frac{a_{\rm max}^2}{T_{\rm min}} \mathcal{A},
\end{equation}
where $T_{\rm min}:= \min_{m,t} [T_{mm}(\theta,t)] \in (0,1)$ is the minimal staying probability, $a_{\rm max}:= \max_{m\ne n, \nu,t}|\partial_{\theta}\ln T_{nm}^{\nu}(\theta,t)|$, and the DA is written explicitly as \cite{lee2024} 
\begin{equation}
    \mathcal{A} := \sum_{i=1}^N \sum_{m\ne n, \nu}P_m(t_{i-1})T_{nm}^{\nu}(\theta,t_{i-1}).
\end{equation}

\subsubsection*{Absolute irreversibility}
We finally consider Markov jump processes involving absolute irreversibility \cite{Murashita2014,Murashita2017} such as the existence of absorbing states. The transitions in these systems can be classified into bidirectional and unidirectional transitions. The transition rates associated with a bidirectional transition satisfy the condition that if $W_{ij}^{\nu,{\rm (b)}} > 0$, then $W_{ji}^{\nu,{\rm (b)}} > 0$ must hold; a unidirectional transition means that if $W_{ij}^{\nu,{\rm (u)}} > 0$, then $W_{ji}^{\nu,{\rm (u)}}$ must vanish. The R-KUR \eqref{eq:bound} is still valid because the DA can now be written as
\begin{equation}
    \mathcal{A}:=\mathcal{A}^{\rm (b)} + \mathcal{A}^{\rm (u)},
\end{equation}
where $\mathcal{A}^{\rm (b/u)}:= \int_0^{\tau}\text{d}t \sum_{i\ne j}W_{ij}^{\rm (b/u)}p_j$ is the traffic due to bidirectional/unidirectional transitions, respectively. We notice that the discrete-time R-KUR \eqref{eq:discrete bound} also applies to absolutely irreversible systems \cite{lee2024}.

\section*{Conclusions}
By establishing a fundamental relationship between the precision of response and the DA, the R-KUR \eqref{eq:bound} paves the way for a deeper understanding of the inherent trade-offs in nonequilibrium processes, particularly in the far-from-equilibrium regime.
The broad applicability of the R-KUR, as demonstrated through examples and various extensions, underscores its potential applications ranging from biochemical processes (e.g., biochemical sensing, switching, and molecular motors) to engineered nanostructures operating away from equilibrium. Furthermore, the ability of the R-KUR to achieve its equality in discrete-state systems highlights its strength and tightness. The R-KUR is expected to serve as a foundation for developing more comprehensive theories of complex systems under nonequilibrium conditions. 


Another feature of our results is that all involved quantities are experimentally measurable. Quantities involved in various results derived in stochastic thermodynamics, such as EP, DA, and the current statistics, can be measured from the statistics of stochastic jumps \cite{Hoang2018,seifert2019,Harunari2022,baiesi2024}.
It is worth mentioning that the measurement of DA can sometimes be more accessible than that of EP, as the latter requires distinguishing the direction of jumps.
Candidates for physical implementation that allow stochastic trajectories to be tracked with a high resolution include a single defect center in diamond \cite{Schuler2005}, single-electron boxes at low temperatures \cite{Koski2013,Koski2014}, molecular motors \cite{Hwang2018}, NMR setups \cite{pal2020}, nanoscale electronic conductors \cite{Friedman2020}, and superconducting NISQ quantum processors \cite{Yu2024}, etc.

Our results are derived for discrete-space Markov processes. It is known that the overdamped Langevin equation can be regarded as the continuous-space limit of the master equation \cite{vandenbroeck2010a}, we therefore expect that the R-KUR \eqref{eq:bound} is also applicable to overdamped Langevin systems. However, although the DA has been studied in previous works \cite{Fullerton2013,Falasco2016}, the connection between the Fisher information and the DA and the explicit formulation of $a_{\rm max}$ remain elusive. It is interesting and important to derive the explicit expression of our bound for overdamped Langevin equation because some experiments in stochastic thermodynamics are implemented by a continuous-space physical setup, such as the thermal rheometer and the colloidal particles \cite{Toyabe2010,Ciliberto2017,Paneru2020}. Finally, for underdamped scenarios, additional modifications in the bound may be necessary due to the non-Markovianity \cite{vanvu2019underdamped}.



Inequality \eqref{eq:bound} demonstrates a fundamental trade-off between sensitivity and precision in stochastic systems. For a fixed number of transitions, high sensitivity implies large fluctuations or low precision. This is apparently similar to the behavior of a thermodynamic system in the vicinity of a critical point where the susceptibility is extremely large and the fluctuations of the order parameter are also significant. On the other hand, due to the intimate relationship between the DA and the nonequilibrium critical phenomena \cite{Garrahan2007,Lecomte2007,Baiesi2009,Fullerton2013},
exploring any potential connections between the R-KUR \eqref{eq:bound} and thermodynamic phase transitions holds promise for intriguing investigations.
Our results may also be applied to the field of thermodynamic inference, which has been attracting increasing interest in recent years \cite{seifert2019,otsubo2020,manikandan2020,vanvu2020estimation,roldan2021,meer2022,Otsubo2022,dieball2023,blom2024}.
A scenario of potential interest involves the estimation of $a_{\max}$ in the absence of detailed model information, with the dynamical activity readily accessible through the counting of transitions.

{
\emph{Note added: we have noticed Ref.~\cite{kwon2024}, which was uploaded to the preprint server subsequent to our submission.
Part of its main results is congruent with applying our R-KUR  to steady states.
This attests to the solidity of our work and indicates that the two studies are mutually complementary.
}
}

\section*{Methods}
\subsection*{Derivation of the R-KUR \eqref{eq:bound}}
The mathematical foundation for deriving our bound is the generalized Cram\'er-Rao inequality \cite{cover1999}.
This inequality addresses scenarios where the probability distribution of a random variable $X$ depends
on a parameter $\theta$, aiming to estimate the function $\psi(\theta)$ of $X$ through an estimator $\Psi(X)$. Hence, $\langle \Psi(X) \rangle_\theta := \int_X {\rm d}X \Psi(X)P_{\theta}(X) = \psi(\theta)$.
The generalized Cram\'er-Rao bound is explicitly given by
\begin{equation}
\label{eq:CR}
	\langle \langle \Psi \rangle \rangle_\theta  \ge \frac{[\text{d}_\theta \psi(\theta)]^2}{I(\theta)}.
\end{equation}
where $I(\theta)$ is the Fisher information, and can be straightforwardly proved with the Cauchy-Schwarz inequality as follows:
\begin{equation}
\label{eq:cs}
    \begin{aligned}
                \langle \langle \Psi \rangle \rangle_\theta I(\theta) & = \int \text{d}X (\Psi - \psi)^2 P_\theta \int \text{d}X (\partial_\theta \ln P_\theta)^2 P_\theta \\
                &\ge \left [\int \text{d}X (\Psi - \psi)(\partial_\theta \ln P_\theta) P_\theta  \right]^2 \\
               & = [\text{d}_\theta \psi(\theta)]^2.
    \end{aligned}
\end{equation}

Now we apply the generalized Cram\'er-Rao inequality to Markov jump processes. We designate $\omega$ as a stochastic trajectory for one realization, 
i.e., $\omega = (i_0, t_0=0; i_1, \nu_1, t_1; i_2, \nu_2, t_2; \dots; i_n, \nu_n, t_n\le \tau =: t_{n+1})$, where a transition occurs from state $i_{m-1}$ to $i_{m}$ through the $\nu_{m}$th channel at time $t_m$,
$P_\theta[\omega]$ as the probability for observing a trajectory $\omega$ with a parameter $\theta$, and $\Psi[X] = O[\omega]$ as an observable defined along the trajectory $\omega$. The path probability takes the form as \cite{Esposito2010}
\begin{strip}
\begin{equation}
    P_\theta[\omega]= p_{i_0}e^{\prod_{m=1}^n e^{\int_{t_{m-1}}^{t_m}\text{d}t W_{i_{m-1}i_{m-1}}(\theta, t) } W_{i_{m}i_{m-1}}^{\nu_m}(\theta,t_m) + \int_{t_n}^{\tau}\text{d}t W_{i_n i_n}(\theta,t) },
\end{equation}
\end{strip}
\noindent where $p_{i_0}$ is the initial distribution. For this path probability, the standard techniques in stochastic thermodynamics yield the Fisher information as a simple form \cite{dechant2019,liu2020,shiraishi2021}
\begin{equation}
\label{eq:fi}
\begin{aligned}
I(\theta)= &
\int_{0}^{\tau}\text{d} t  \sum_{i\ne j, \nu} W_{ij}^\nu p_j (\partial_\theta \ln W_{ij}^\nu  )^2	\\
& \le a_{\max}^2 \mathcal{A},  
\end{aligned}
\end{equation}
where 
\begin{equation}
    a_{\max} := \mathop{\max_{ 
i\ne j, \nu}}_{t\in [0,\tau]}
|\partial_\theta \ln W_{ij}^\nu(\theta,t)|,
\end{equation}
and $\mathcal{A}$ is the time-integrated DA defined as
\begin{equation}
	\mathcal{A} :=  \int_0^\tau \text{d}t  \sum_{i\ne j} W_{ij}(\theta,t)p_j(t)	.
\end{equation}
The DA $\mathcal{A}$ is the ensemble average of the total number of transitions that occurred within the time interval, which is written explicitly as
\begin{equation}
    A_{\rm e}[\omega] = \sum_{i\ne j, \nu}n_{ij}^{\nu}[\omega] = \sum_{i\ne j, \nu}\sum_{m=1}^{n} \delta_{i_{m-1},j}\delta_{i_{m},i}\delta_{\nu_{m},\nu},
\end{equation}
where $n_{ij}^{\nu}[\omega]:=\sum_{m=1}^{n} \delta_{i_{m-1},j}\delta_{i_{m},i}\delta_{\nu_{m},\nu}$ counts the total number of transitions from state $j$ to state $i$ via channel $\nu$.

Combining the explicit expression of the Fisher information with the Cram\'er-Rao inequality \eqref{eq:CR} yields the desired inequality \eqref{eq:bound}. Because the derivation does not impose any constraint on the form of the observable $O$, it is not necessary to be a current or an absolute counting variable.

\subsection*{Details of the calculations in Example 1}
The empirical accumulated current for a time interval $\tau$ is $J_{\rm e}[\omega] = n_+ - n_-$, where $n_{+/-}$ represents the total number of transitions that the particle jumps to the positive/negative direction. In the NESS, the ensemble-averaged value is \cite{barato2015,hiura2021,vo2022}
\begin{equation}
    J = \tau e^{\theta} (e^{\frac{\alpha}{2}} - e^{-\frac{\alpha}{2}}).
\end{equation}
The response of the current to a perturbation in $\theta$ or $\alpha$ is, respectively,
\begin{align}
    & \text{d}_{\theta} {J}(\theta) = {J}(\theta);\\
    & \text{d}_{\alpha} {J}(\alpha) = \frac{\tau e^{\theta}}{2} (e^{\frac{\alpha}{2}} + e^{-\frac{\alpha}{2}}).
\end{align}
We also have $a_{\rm max} = 1$ for a perturbation in $\theta$ and $a_{\rm max} = 1/2$ for $\alpha$, respectively. The variance, the total EP and the DA are
\begin{align}
    & \langle \langle {J} \rangle \rangle = \tau e^{\theta} (e^{\frac{\alpha}{2}} + e^{-\frac{\alpha}{2}});\\
    & \sigma = \tau e^{\theta} \alpha (e^{\frac{\alpha}{2}} - e^{-\frac{\alpha}{2}});\\
    & \mathcal{A} = \tau e^{\theta} (e^{\frac{\alpha}{2}} + e^{-\frac{\alpha}{2}}).
\end{align}

\subsection*{Details of FCS in Example 2}
FCS is a powerful tool to calculate the moments and the cumulants of a random variable. The central quantity of FCS is the cumulant generating function \cite{bagrets2003a,ren2010,gu2018,liu2020,gu2023}
\begin{equation}
    {G}_\chi (\tau) = \ln \big[\mathbf{1} \mathbb{T} e^{\int_0^\tau \mathcal{W}_\chi(t) \text{d} t} \mathbf{p}(0) \big],
\end{equation}
where $\mathbf{1}=[1,1]$, $\mathbb{T}$ is the time-ordering operator, $\mathbf{p}(0)$ is the initial probability distribution, and
\begin{equation}
    \mathcal{W}_\chi(t) = 
    \begin{pmatrix}
        - W_{10}^{\rm h} -W_{10}^{\rm c} &  W_{01}^{\rm h} + W_{01}^{\rm c} e^{\chi}    \\ 
        W_{10}^{\rm h} + W_{10}^{\rm c} e^{-\chi} & -W_{01}^{\rm h} - W_{01}^{\rm c} 
    \end{pmatrix}
\end{equation}
is the transition rate matrix with a counting field $\chi$ for the first observable, i.e., the empirical accumulated current that flows into the cold bath $J^{\rm c}_{\rm e}[\omega]:= \sum_{i\ne j}(n_{ij}^{\rm c}[\omega] - n_{ji}^{\rm c}[\omega])$. 
Its average and variance
are calculated by the derivatives of the cumulant generating function as
\begin{equation}
        J^{\rm c} = \left. \frac{\partial G_\chi(\tau)}{\partial \chi} \right|_{\chi = 0}, \quad \langle \langle J^{\rm c} \rangle \rangle = \left. \frac{\partial^2 G_\chi(\tau)}{\partial \chi^2} \right|_{\chi = 0}.
\end{equation}

The second observable we simulated is the empirical time-integrated traffic due to coupling with the cold bath, defined as $A_{\rm e}^{\rm c}[\omega] := \sum_{i\ne j} n_{ij}^{\rm c}[\omega]$. The ensemble average and the variance can also be calculated via FCS by introducing a different counting field $\lambda$ as \cite{Gopich2006}
\begin{equation}
    {G}_\lambda (\tau) = \ln \big[\mathbf{1} \mathbb{T} e^{\int_0^\tau \mathcal{W}_\lambda(t) \text{d} t} \mathbf{p}(0) \big],
\end{equation}
where
\begin{equation}
    \mathcal{W}_\lambda(t) = 
    \begin{pmatrix}
        - W_{10}^{\rm h} -W_{10}^{\rm c} &  W_{01}^{\rm h} + W_{01}^{\rm c} e^{\lambda}    \\  
        W_{10}^{\rm h} + W_{10}^{\rm c} e^{\lambda} & -W_{01}^{\rm h} - W_{01}^{\rm c} 
    \end{pmatrix}.
\end{equation}

In the numerical simulation of both observables, we sample each of the parameters contained in the energy gap $\Delta$ uniformly from the following intervals: $\beta_{\rm c} \in [1,5]$, $\varepsilon \in [0.1,1]$, $C \in [0,2]$, $s\in [-0.5,0.5]$, $a \in [0,1]$, $\omega \in [0,5]$, and $\tau \in [0,1]$.

\subsection*{Derivation of the R-KUR for feedback-controlled systems}
The Fisher information for a feedback-controlled process takes the following form \cite{liu2020}
{\small
\begin{equation}
    I(\theta) = \int_0^{\tau}\text{d}t \sum_{i\ne j, \nu, m}W_{im,jm}^{\nu}(\theta,t)p_{jm}(t)(\partial_{\theta}\ln W_{im,jm}^{\nu}(\theta,t))^2,
\end{equation}}
where $p_{jm}$ is the joint probability of $\mathbb{S}$ being in state $j$ and $\mathbb{M}$ giving an outcome $m$, and $W_{im,jm}^{\nu}$ is the transition rate from the joint state $jm$ to state $im$ via channel $\nu$. Therefore, we obtain an upper bound
\begin{equation}
    I(\theta) \le \sum_m p_m a_{m,\rm max}^2 \mathcal{A}_m \le a_{\rm max}^2 \mathcal{A},
\end{equation}
where $a_{m,\rm max}:= \max_{ 
i\ne j, \nu,t\in [0,\tau]}|\partial_{\theta}\ln W_{im,jm}^{\nu}(\theta,t)|$ is optimized for each outcome $m$, $a_{\rm max} := \max_{m}[a_{m,\rm max}]$ is in turn optimized against all possible outcome $m$'s, the conditional DA for an outcome $m$ is defined as
\begin{equation}
    \mathcal{A}_m = \int_0^{\tau} \text{d}t \sum_{i\ne j, \nu} W_{im,jm}^{\nu}(\theta, t)p_{j|m}(t),
\end{equation}
and the total DA is then
\begin{equation}
    \mathcal{A} = \sum_m p_m\mathcal{A}_m.
\end{equation}
We notice that the probability $p_m$ of obtaining an outcome $m$ is independent of time because the state of $\mathbb{M}$ does not evolve inside one feedback loop.

\subsection*{Derivation of the R-KUR for discrete-time Markov chains \eqref{eq:discrete bound}}
For discrete-time Markov chains, according to Ref.~\cite{liu2020}, the Fisher information takes the form as
{\small
\begin{align}
    I(\theta) & = -\sum_{i=1}^N \sum_{m,n,\nu} P_m(t_{i-1})T_{nm}^{\nu}(\theta,t_{i-1})\partial^2_{\theta}\ln T_{nm}^{\nu}(\theta,t_{i-1}) \nonumber\\
    & =: I_{\rm od}(\theta) + I_{\rm d}(\theta),
\end{align}}where the off-diagonal term of the Fisher information $I_{\rm od}(\theta)$ and the diagonal term $I_{\rm d}(\theta)$ are respectively defined as
{\small
\begin{align}
    & I_{\rm od}(\theta) := \sum_{i=1}^N \sum_{m\ne n,\nu}P_m(t_{i-1})T_{nm}^{\nu}(\theta,t_{i-1})(\partial_{\theta}\ln T_{nm}^{\nu}(\theta,t_{i-1}))^2; \\
    & I_{\rm d}(\theta) := \sum_{i=1}^N \sum_{m,\nu}P_m(t_{i-1})T_{mm}(\theta,t_{i-1})(\partial_{\theta}\ln T_{mm}(\theta,t_{i-1}))^2.
\end{align}}

We now define the DA of a discrete-time Markov chain. By definition, the DA quantifies the time scale of  transitions. From time $t_{i-1}$ to time $t_i$, the probability of a transition occurring between state $m$ and state $n$ through channel $\nu$ is given by
\begin{equation}
    a_{mn}^{\nu}(t_{i-1}) := P_m(t_{i-1})T_{nm}^{\nu}(\theta,t_{i-1}) + P_n(t_{i-1})T_{mn}^{\nu}(\theta,t_{i-1}).
\end{equation}
Therefore, the DA is
\begin{align}
    \mathcal{A} &:= \frac{1}{2}\sum_{i=1}^N \sum_{m\ne n, \nu} a_{mn}^{\nu}(t_{i-1}) \nonumber\\
    & = \sum_{i=1}^N \sum_{m\ne n, \nu}P_m(t_{i-1})T_{nm}^{\nu}(\theta,t_{i-1}).
\end{align}
This definition can also be found in Ref. \cite{lee2024}.

We can prove that the Fisher information, its diagonal part, and the off-diagonal term are all bounded from above by the DA. It is obvious that the off-diagonal term of the Fisher information is bounded by the DA as
\begin{equation}
    I_{\rm od}(\theta) \le a_{\rm max}^2 \mathcal{A},
\end{equation}
where $a_{\rm max}:= \max_{m\ne n, \nu,t}|\partial_{\theta}\ln T_{nm}^{\nu}(\theta,t)|$.

The diagonal term is bounded as
\begin{strip}
    \begin{align}
    I_{\rm d}(\theta) & = \sum_{i=1}^N \sum_{m}P_m(t_{i-1}) \frac{(\sum_{n:n\ne m, \nu} \partial_{\theta} T_{nm}^{\nu}(\theta,t_{i-1}))^2}{T_{mm}(\theta,t_{i-1})}\nonumber\\
    & = \sum_{i=1}^N \sum_{m}P_m(t_{i-1}) \frac{(\sum_{n:n\ne m, \nu} \partial_{\theta} T_{nm}^{\nu}(\theta,t_{i-1}))^2}{\sum_{n:n\ne m, \nu} \partial_{\theta} T_{nm}^{\nu}(\theta,t_{i-1})} \frac{1- T_{mm}(\theta,t_{i-1})}{T_{mm}(\theta,t_{i-1})} \nonumber\\
    & \le \sum_{i=1}^N \sum_{m}P_m(t_{i-1}) \sum_{n:n\ne m, \nu} \frac{(\partial_{\theta} T_{nm}^{\nu}(\theta,t_{i-1}))^2}{\partial_{\theta} T_{nm}^{\nu}(\theta,t_{i-1})} \frac{1- T_{mm}(\theta,t_{i-1})}{T_{mm}(\theta,t_{i-1})} \nonumber\\
    & = \sum_{i=1}^N \sum_{m\ne n, \nu}P_m(t_{i-1}) T_{nm}^{\nu}(\theta,t_{i-1}) (\partial_{\theta} T_{nm}^{\nu}(\theta,t_{i-1}))^2 \frac{1- T_{mm}(\theta,t_{i-1})}{T_{mm}(\theta,t_{i-1})} \nonumber\\
    & \le \frac{1- T_{\rm min}}{T_{\rm min}} I_{\rm od}(\theta) \nonumber\\
    & \le \frac{1- T_{\rm min}}{T_{\rm min}}a_{\rm max}^2 \mathcal{A},
\end{align}
\end{strip}
\noindent where $T_{\rm min}:= \min_{m,t} [T_{mm}(\theta,t)]$ is the minimal staying probability and we use Titu's lemma in the first inequality.

Finally, combining all the inequalities yields
\begin{equation}
    I(\theta)\le \frac{a_{\rm max}^2}{T_{\rm min}} \mathcal{A}.
\end{equation}

\section*{Data availability}
All data needed to supporting the conclusions of this study are present in the paper.

\section*{Code availability}
All the computational codes that were used to generate the data presented in this study are available from the corresponding authors upon reasonable request.
\bibliographystyle{unsrt}
\bibliography{ref}

\section*{Acknowledgements}
The authors appreciate Prof. Ying Tang for fruitful discussions. This work was supported by the Scientific Research Start-up Foundation of Xihua University (Grant No: Z241064).

\section*{Author contributions}
K.L. proved the main result, conducted the first example and generalized the results.
J.G. initiated the project and conducted the second example.
Both authors discussed the findings, drew the conclusions and wrote the paper.
\section*{Competing interests}
The authors declare no competing interests.
\end{document}